\documentclass[12pt]{article}
\title{The problem of quantization of lightcone QCD}

\author{Alexey V. Popov\thanks{email: avp@novgorod.net} \\ \\ 
\small{\emph{Velikiy Novgorod, Russia}} 
}
\date{}

\begin{document}
\maketitle
\abstract{
There exists the problem to construct a quantum algebra of observables in lightcone QCD beyond the perturbative regime.
It has recently established that the boundary gauge fields are crucial for a consistent construction of the classical dynamic system.
If the gauge group is non-Abelian and there are four or more space-time dimensions 
then the procedure of symplectic reduction gives a classical dynamical system with very complicated Hamiltonian having infinite power over the coupling constant.
Then, to quantize the theory one should to construct a Poisson algebra and to quantize it. 
Careful analysis shows that a Poisson formulation has a problem with:
canonical commutation relations, spatial invariance, and the boundary degrees of freedom in the Hamiltonian.
}
\newcommand{\ket}[1]{\ensuremath{|#1\rangle}}
\newcommand{\bra}[1]{\ensuremath{\langle#1|}}
\newcommand{\braket}[2]{\ensuremath{\langle#1|#2\rangle}}
\section{Introduction}
It was shown recently that in lightcone QCD the boundary gauge fields at $x^-=\pm\infty$ are important part of the theory \cite{avp_Hqcd,Kovner07,Zhang1}.
Since there are no natural boundary conditions in QCD, we must carefully treat boundary fields during a construction of the quantum theory. 
The practical importance of the $x^-$-boundaries was established in Ref. \cite{Kovner07} by the considering of the theory of high energy evolution.
However, the theoretic input, used in Ref. \cite{Kovner07}, is not fully verified by a rigorous formalism.
The construction of a such formalism had been started in Ref. \cite{avp_Hqcd}, where the classical lightcone QCD is considered 
but the problem of quantization is not examined in detail.
Hoping to stimulate future researches, in this paper we analyze the currently known attempts to quantize the theory and show why all they fail.

Refs. \cite{Zhang1,Zhang2} is the first known papers where the role of a field asymptotic was established.
However, in those papers only the antisymmetric boundary condition was used. 
Nowadays, we know that in QCD the boundary condition is neither antisymmetric nor symmetric.
Next, in Ref. \cite{Kovner07} the boundary fields is used in the theory of high energy scattering, 
since the method of wave function requires the precisely defined quantum Hamiltonian. 
The complete description of the lightcone QCD at the classical level was given in Ref. \cite{avp_Hqcd}.
Nevertheless, a quantization procedure of the obtained system is not specified yet.
The key problem here is to construct a corresponding Poisson algebra, which, however, is unable to treat boundary fields. 
A Poisson algebra is required, since this is the standard way of quantization. 
Finally, a Poisson bracket should be quantized and a representation of the quantized algebra in a Hilbert space should be found.
For a theory without boundary fields this procedure is simple enough. 
However, even for a slightly more complex theory, the procedure may become unsolvable.
In this paper we show that in lightcone QCD a Poisson algebra cannot be directly constructed, 
since the Hamiltonian has both the explicit and the implicit boundary contributions.
Hence, we have to find a workaround that can help to construct an algebra.
We analyze the possible choices of variables, in which a required Poisson algebra is expected to be constructed.
The result is negative, the standard method of quantization cannot be applied.

This paper is organized as follows.
In Sec. \ref{sect_1}, we briefly review lightcone QCD at the classical level.
In Sec. \ref{sect_2}, we analyze the available choices of coordinates with aim to construct a Poisson algebra and to quantize it.
Section \ref{sect_3} contains our conclusions.

\section{Lightcone QCD at the classical level} \label{sect_1}
Let us briefly review the results of Ref. \cite{avp_Hqcd}. The Lagrangian of QCD is
\begin{equation}
L=\frac{1}{2}{F_{+-}}^2+F_{+i}F_{-i}-\frac{1}{4}F_{ij}F^{ij}+gA_\mu J^\mu
\end{equation}
where $J^\mu$ is a current of matter fields or an external current.
To remove unphysical degrees of freedom we apply the symplectic Faddeev-Jackiw method \cite{Faddeev}.
This method allows to do the job without a construction of a Poisson algebra, which has a problem with boundary degree of freedom. 

The variables $A_+$ are automatically removed from the scene during the process of symplectic reduction, 
since there is no term $\partial_+A_+$ in the Lagrangian.
The main gauge fixing is
\begin{equation}
A_-=0
\end{equation} 
The necessity to reduce the Hamiltonian on the phase space without $A_+$ gives the complete set of Gauss' constraints.
\begin{equation} \label{eq_3}
\partial_-\pi^-_a + \partial_- \partial_i A_i^a-gf_{abc}\partial_-A_i^bA^c_i+gJ^+_a=0
\end{equation}
\begin{equation} \label{eq_4}
\pi^-_a(\pm\infty,\vec x)=0
\end{equation}
where the last constraint comes from the boundary part of the variation.

After the removal of unphysical degrees of freedom we obtain the following classical dynamical system.
The phase space of the theory is the space of fields $A_i^a(x^-,\vec x)$ obeying the boundary condition
\begin{equation} \label{eq_7}
A^a_i(-\infty,\vec x)=0
\end{equation}
This condition was selected in \cite{avp_Hqcd} as a most simple way to fix the residual gauge freedom.
The linear space of fields $A_i^a$ is endowed by the fundamental  symplectic form $\omega$
\begin{equation}
\omega= \int \partial_- d A_i^a \wedge dA_i^a
\end{equation} 
\begin{equation} \label{eq_5}
\omega(A,B)= \int (\partial_- A_i^a  B_i^a - A_i^a \partial_- B_i^a ) dx^-
\end{equation}
The Hamiltonian density is given by
\begin{equation} \label{eq_10}
H[A_i]=\frac{1}{2} (\pi^-_a)^2+\frac{1}{4}F_{ij}^a[A_i] F^{ij}_a[A_i]  - g A_i^a J^i_a
\end{equation}
The momentum $\pi^-_a(x^-,\vec x)$ is determined by the equations (\ref{eq_3}) and (\ref{eq_4}), which is the Gauss' constraints.
At the boundary $x^-=+\infty$ we have the constraint
\begin{equation}
A^a_i(+\infty,\vec x)=\gamma_i^a(\vec x)
\end{equation}
\begin{equation} \label{eq_1}
\partial_i \gamma_i^a(\vec x)= \int\limits_{-\infty}^{+\infty} \left(gf_{abc}\partial_- A_i^b  A^c_i-gJ^+_a\right) dx^-
\end{equation}
which is the consequence of the Gauss' constraints (\ref{eq_3}) and (\ref{eq_4}). We also add the following constraint
\begin{equation} \label{eq_2}
\partial_i  \gamma_j^a-\partial_j  \gamma_i^a+g f_{abc} \gamma^b_i \gamma^c_j=0
\end{equation}
which is the finite energy condition for infinity $x^-$-dimension and introduced manually to complete fix $\gamma_i^a$.
Formally, the constraint (\ref{eq_2}) does not arise during the symplectic Faddeev-Jackiw algorithm. 
So, it must be used carefully, since in a quantum theory energy levels have a sense only relatively to a ground state.

\section{The problem of quantization} \label{sect_2}
Initially, we have a symplectic system.  
Note, that in the classical mechanic the symplectic form $dp\wedge dq$ is a natural part of the construction of the Hamiltonian formalism.
Next, the Poisson bracket $\{q,p\}=1$ is obtained by the implicitly inversion on the symplectic form.
Usually, this step is not emphasized because it is trivial.
However, subtleties may emerge in an infinite-dimensional field theory and in a constrained system.

In a quantum theory a role of Poisson algebra is much more important because a quantum commutator is a quantized version of the classical brackets.
The mainstream method of quantization is
\begin{enumerate}
\item To construct a Poisson algebra with a Hamiltonian. This algebra will be used as an algebra of classical observables. 
\item To fix a gauge invariance and to remove constraints by a construction of the Dirac bracket. 
\item To quantize the Poisson algebra. To construct a Hilbert space and canonical commutation relations.
\item To resolve possible ordering ambiguities in quantum-classical operator correspondence.
\end{enumerate}
The second step we perform within symplectic system. 
To make the first step we analyze the three variables with aim to generate well defined and quantizable Poisson algebra.

\subsection{Natural variables $A_i^a$} \label{subsect_2_1}
To construct a Poisson algebra from our symplectic dynamical system we have to invert the symplectic form (\ref{eq_5}) and to construct a Poisson brackets.
Let us try to invert the symplectic form in the linear space of fields obeying the boundary condition $A^a_i(-\infty,\vec x)=0$.
We have to find a linear operator $P(x-y)$ such that 
\begin{equation} \label{eq_6}
\omega\left(A,\int P(x-y) B(y) dy\right)=\int A B dx
\end{equation} 
for any $A$ and $B$ obeying (\ref{eq_7}). Note that we have omit the color and transverse indexes. 
Then, the corresponding Poisson brackets is
\begin{equation} \label{eq_11}
\{F,G\}=\int \frac{\delta F[A]}{\delta A(x)} P(x-y)  \frac{\delta G[A]}{\delta A(y)} dx dy
\end{equation}
It is easy to check that for the boundary condition (\ref{eq_7}) there exists only one such operator
\begin{equation} \label{eq_8}
P(x) = \frac{1}{2}\theta (-x)
\end{equation}
where $\theta (x)$ is the standard step function.
The result (\ref{eq_8}) has a fatal problem: the kernel $P(x)$ is not antisymmetric. 
This obstruction is expected specific feature of a infinite-dimensional theory with boundaries. 
Hence, a naive Poisson brackets does not exist and the variables $A_i^a$ cannot generate a Poisson algebra of observables.
In principle, this way is not fully closed and it may be exist a subtle method to construct an algebra with a Hamiltonian, but we don't know it now.

\subsection{Antisymmetric variables $\tilde A_i^a$} \label{subsect_2_2}
To overcome the problem we have to try to find an other set of canonical variables that can give a Poisson brackets.
Indeed, from the wide practice of lightcone field theories we know that antisymmetric or zero boundary conditions induces 
a Poisson algebra with the lightcone canonical Poisson brackets. 
For the such boundary conditions the symplectic form (\ref{eq_5}) is invertible \cite{avp_Hqcd} and the corresponding inverse kernel is
\begin{equation} \label{eq_9}
P(x)=-\frac{1}{4} \varepsilon(x)
\end{equation}
where $\varepsilon(x)$ is the sign function. The kernel (\ref{eq_9}) gives the canonical lightcone Poisson brackets
\begin{equation} \label{eq_19}
\{\tilde A_i^a(x^-,\vec x),\tilde A_j^b(y^-,\vec y)\}= - \frac{1}{4} \varepsilon(x^--y^-) \delta(\vec x - \vec y) \delta_{ij} \delta^{ab}
\end{equation}
How to use the canonical result (\ref{eq_9}) in our dynamical system? 
In Ref. \cite{avp_Hqcd} we prove that in lightcone QCD the antisymmetric boundary conditions cannot be achieved by a residual gauge transformation.
It was proposed in Ref. \cite{Kovner07} to introduce new variables with antisymmetric boundary conditions 
\begin{equation}
\tilde A = A-\frac{1}{2}\gamma
\end{equation}
It was shown in \cite{avp_Hqcd} that such transformations are canonical. 
However, a potential problem here is in a redefinition of a Hamiltonian in a quantum theory because it must be redefined in terms of new variables $\tilde A$.
It is clear that in a quantum theory a corresponding operator $\hat \gamma$ cannot be simple defined, since it is a boundary degree of freedom.
In our case of lightcone QCD we have a chance to define it due to constraints (\ref{eq_1}) and (\ref{eq_2}), 
which completely fix $\gamma$ and allow to express it in terms of the bulk degrees of freedom $\tilde A$. 
At first sight, it is sufficient to define a new Hamiltonian as
\begin{equation}
\tilde H[\tilde A]=H[\tilde A + \frac{1}{2}\gamma[\tilde A]]
\end{equation} 
where the functional $\gamma[\tilde A]$ is a solution of the constraint (\ref{eq_1}) and (\ref{eq_2}).
However, the fatal problem here is that the lightcone QCD Hamiltonian (\ref{eq_10}) has an implicit dependence over the boundary field $\gamma$. 
Indeed, since the Hamiltonian (\ref{eq_10}) is an integral functional, 
it may have a contribution from the boundary by the terms having the derivative $\partial_-$.
If so, the Hamiltonian is not an element of the bulk Poisson algebra and the standard method of quantization can not be applied.
An other interpretation of this situation is that the brackets (\ref{eq_11}) used for $\{H,\tilde A\}$ gives a wrong result, 
since the variational derivative in (\ref{eq_11}) has a boundary term.

To extract the implicit boundary part from the Hamiltonian (\ref{eq_10}) we calculate an infinitesimal variation $\delta \tilde H$  
over a variation of the canonical variables $\delta \tilde A$.
At first, it is necessary to calculate the variation $\delta \gamma$. 
From the constraint (\ref{eq_1}) we have
\begin{equation}
\delta \partial_i \gamma_i^a(\vec x)= \delta \int\limits_{-\infty}^{+\infty} \left(gf_{abc}\partial_- \tilde A_i^b  \tilde A^c_i\right) dx^-
\end{equation}
The boundary part of this variation is proportional to
\begin{equation} \label{eq_12}
\int\limits_{-\infty}^{+\infty} f_{abc}\partial_- ( \delta \tilde A_i^b  \tilde A^c_i) dx^- 
=\left.f_{abc}\delta \tilde A_i^b  \tilde A^c_i \right|_{-\infty}^{+\infty}
\end{equation}
Since both $\tilde A$ and $\delta \tilde A$ obey the antisymmetric boundary conditions, 
the term $\delta \tilde A_i^b  \tilde A^c_i$ obeys symmetric one. So, the boundary term (\ref{eq_12}) is exactly zero and
the variation $\delta \gamma[\tilde A]$ have no boundary terms.
This conclusion shows that the boundary contribution to the variation $\delta \tilde H$ can come only from the term $(\pi^-_a)^2$.

Consider the variation $\delta \pi^-_a(x^-,\vec x)$, which is more complex. 
Rewriting the definition (\ref{eq_3}) in the terms of $\tilde A$ and making the variation we have
\begin{equation} \label{eq_13}
\partial_- \delta \pi^-_a + \partial_- \partial_i \delta \tilde A_i^a-gf_{abc}\delta\left(\partial_-\tilde A_i^b(\tilde A^c_i+\frac{1}{2}\gamma^c_i)\right)=0
\end{equation}
Due to the constraint (\ref{eq_4}) there exist several methods to invert the operator $\partial_-$ but all they give same final result.
As it was used in Ref. \cite{avp_Hqcd} we choose the operator having the kernel
\begin{equation} \label{eq_14}
\partial_-^{-1}(x^--y^-) =\frac{1}{2}\varepsilon(x^--y^-) 
\end{equation} 
Applying $\partial_-^{-1}$ to Eq. (\ref{eq_13}) we see that
the boundary contribution arises only from the term
\begin{equation}
gf_{abc}\partial_-^{-1}(\partial_-\delta\tilde A_i^b) \tilde A^c_i
\end{equation}
Using Leibnitz rule we move $\partial_-$ to the non-variational term 
\begin{equation} \label{eq_15}
gf_{abc}\partial_-^{-1}\left(\partial_-(\delta\tilde A_i^b \tilde A^c_i) -  \delta\tilde A_i^b \partial_-\tilde A^c_i\right)
\end{equation}
The boundary contribution lies within the term $\partial_-^{-1}\partial_-(\delta\tilde A_i^b \tilde A^c_i)$.
Using the definition (\ref{eq_14}), it is easy to show that for any function $f(x^-)$
\begin{equation} \label{eq_16}
\partial_-^{-1}\partial_- f = f(x^-)-\frac{1}{2}\left(f(+\infty)+f(-\infty)\right)
\end{equation}
Note that this also shows that if $f(x^-)$ obeys the antisymmetric boundary condition then $\partial_-^{-1}\partial_- f=f$.
Returning to Eq. (\ref{eq_15}) and applying the identity (\ref{eq_16}), we see that the boundary contribution is
\begin{equation}
\left.\delta \pi^-_a\right|_{\mbox{\scriptsize boundary}}=-\frac{g}{2}f_{abc}\left.\delta\tilde A_i^b \tilde A^c_i \right|_{-\infty}^{+\infty}
=-\frac{g}{4}f_{abc} \delta \gamma_i^b \gamma^c_i
\end{equation}
Finally, the boundary contribution to the variation $\delta \tilde H$ is 
\begin{equation} \label{eq_17}
\left.\delta \tilde H \right|_{\mbox{\scriptsize boundary}}=
-\int\limits_{x^-,\vec x}\frac{g}{4}f_{abc} \pi^-_a \delta \gamma_i^b \gamma^c_i 
\end{equation}
where the functionals $\gamma [\tilde A]$ and $\pi^-_a[\tilde A]$ are assumed.
The bulk part of the variation $\delta \tilde H$ can be calculated in the standard way 
by the assumtion that $\delta \tilde A_i^a$ vanish at the boundary.    

If a theory has boundary part of a variation equal to zero, then the Hamiltonian is an element of standard bulk Poisson algebra of observables
and this theory can be directly quantized without complications.
In the case of four-dimensional QCD the variation (\ref{eq_17}) is not zero.
Hence, the naive quantization is not possible and we have to find an other way.
The most straightforward idea is to convert in a some way the boundary contribution (\ref{eq_17}) to a bulk one by an inversion of the variation.
Namely, let us find a bulk functional $H'[\tilde A]$ such that
\begin{equation}
\delta H'[\tilde A]=\left.\delta \tilde H \right|_{\mbox{\scriptsize boundary}}
\end{equation}
If such $H'$ exists, then the bulk variation of the new Hamiltonian $\tilde H + H'$ exactly restores the boundary variation (\ref{eq_17}).
This indeed may be possible in a scalar theory. Unfortunately, in QCD such $H'$ does not exist.
The most obvious obstruction is that the color-space 1-form $f_{abc} d \gamma^b \gamma^c$ is not exact and not closed,
except the case of a one-dimensional color space.
It should be noted, that perturbative computations do not affected by this problem at least at the leading orders, 
since the variation (\ref{eq_17}) has order $g^3$.
There is no doubt that a quantization of the full theory beyond the perturbative regime must have a solution of this problem.

\subsection{Boundaryless variables $c_i^a$} \label{subsect_2_3}
The another attempt to separate the boundary contribution is proposed in Ref. \cite{Kovner07}. 
Since it was used for the theory of high energy evolution, we shall consider this method in a detailed way.
The idea is simple enough: Let us define a new fundamental variables $c_i^a$ with zero boundary conditions as
\begin{equation} \label{eq_18}
A_i^a(x^-,\vec x) = c_i^a(x^-,\vec x) + \gamma_i^a(\vec x) \varphi(x^-)
\end{equation}
where $\varphi(x^-)$ is an arbitrary fixed global function such that
\begin{equation} 
\begin{array}{l}
\varphi(-\infty)=0 \\
\varphi(+\infty)=1
\end{array}
\end{equation}
where it is helpful to imagine $\varphi(x^-)$ as a typical smooth monotonic kink-like function.
So, in the definition (\ref{eq_18}) we have  $c_i^a(\pm\infty)=0$.
Taking $c_i^a$ as a fundamental variables of the theory, it is necessary to redefine the Hamiltonian (\ref{eq_10}) and $\gamma$ in terms of the new variables $c_i^a$.
This gives the new Hamiltonian $H[c]$, which variation $\delta H[c]$ is well defined due to the constraint $\delta c(\pm\infty)=0$.
While the Hamiltonian can be redefined easily, a determination of $\gamma[c]$ from the equation (\ref{eq_1}) becomes very complicated and
nonlinear. We only assume here that such $\gamma[c]$ exists.

The relation (\ref{eq_18}) can be viewed as a linear coordinate map $A\to c$, 
since $\gamma=A(+\infty)$ by the definition.
After the imposition of the constraint $A(+\infty)=\gamma[A]$, we obtain a constraint manifold and the map becomes nonlinear.
The inverse map $c \to A$ is also strongly nonlinear due to inherent nonlinearity of the functional $\gamma [c]$.
Hence, the linear set of variables $c_i^a$ cannot be endowed by the induced symplectic form 
$\omega_c(c_1,c_2)=\omega(A_1[c_1],A_2[c_2])$.
So, the variables $c_i^a$ are not symplectic and they can have a sense only at the stage of Poisson algebra, which is still not constructed.
Note that, before an imposition of the constrain the map $A \to \tilde A$ and its inverse $\tilde A \to A$ are linear maps as it was argued in Ref. \cite{avp_Hqcd}.
Hence the linear set of the fields $\tilde A$ is a linear symplectic space.

It is possible to construct a Poisson algebra generated by the variables $c_i^a$.
Using the nonlinear inverse map $c \to A$, we express the Hamiltonian (\ref{eq_10}) in terms of the variables $c_i^a$
\begin{equation}
H_c[c]=H[A[c]]=H[c+\gamma[c]\varphi]
\end{equation}
Now the variation $\delta H_c$ over $\delta c$ is well defined, since $c_i^a$ obey zero boundary condition.
Hence, $H_c[c]$ can play a role of a Hamiltonian.

A failure of this method is that the fundamental brackets $\{c(x_1),c(x_2)\}$ becomes nonlinear.
Using the nonlinear map $\tilde A \to c$
\begin{equation} 
c[\tilde A]= \tilde A - \gamma [\tilde A] \left(\varphi-\frac{1}{2}\right)
\end{equation}
and the fundamental brackets (\ref{eq_19}) for $\tilde A$, we have
\begin{equation}  \label{eq_20}
\{c(x_1)[\tilde A],c(x_2)[\tilde A]\}=B(x_1,x_2)[\tilde A]
\end{equation}
where $B(x_1,x_2)[A]$ is a some functional which precise form is not important here.
The brackets (\ref{eq_20}) is well defined, since the variation $\delta c[\tilde A]$ 
has only a bulk contribution because it is proved above that $\delta \gamma[\tilde A]$ is a bulk functional.
Replacing in the functional $B[\tilde A]$ the variables $\tilde A$ by $\tilde A[c]$
\begin{equation}
\tilde A[c]=c+\gamma[c]\left(\varphi-\frac{1}{2}\right)
\end{equation}
we prove that the variables $c_i^a$ generate a closed Poisson algebra with the Hamiltonian $H_c[c]$.
This algebra is an Poisson algebra of functionals over $c_i^a$. 
Also, it is a Poisson subalgebra of the original algebra generated by $\tilde A$.
There is no idea how to quantize the nonlinear brackets (\ref{eq_20}).
Though, on the classical level the variables $c_i^a$ may be useful \cite{Kovner07}.

The usage of variables $c_i^a$ has the one more problem. 
Since the function $\varphi(x^-)$ explicitly depends on $x^-$, it breaks the longitudinal spatial invariance of the Hamiltonian.
This problem is most serious because there are no preferred points in the initial formulation of the theory.
Although this does not change a physical context, the math formulation seems ugly. 

\section{Discussion} \label{sect_3}
We have analyzed the three possible variables: $A$, $\tilde A$, and $c$. 
All they give the fatal obstructions that do not allow to quantize the theory by the standard way.  
This motivates us to find a different quantization scheme not based on a bulk Poisson algebra.
In Refs. \cite{Soloviev,Bering} it was proposed a generalization of Poisson brackets to a case with boundaries.
It is possible to construct a consistent Poisson bracket that includes the full boundary information. 
For boundaryless functionals a such bracket coincides with the standard Poisson brackets, 
but for functionals, with a boundary contribution to a variation, a bracket is defined by a distinct way.
Although this generalized bracket seem challenging, there are no hopes to quantize such brackets in a trivial way.
Moreover, it was shown that a such bracket is not unique \cite{Bering,Soloviev99}. 
It is not clear what bracket naturally corresponds to a symplectic structure obtained from a Lagrangian. 
The task becomes more complex if we impose the constraint $\tilde A(\pm\infty)=\pm\gamma/2$ that must be treated as a second class constriant. 

Another quantization method, which we did not touch before, is an introduction of unphysical degrees of freedom into a Hilbert space of the 
quantum theory. Sometimes, such methods leads to unitary ghosts and Faddeev-Popov ghosts in a Hilbert space.
A determination of physical states is given by operator conditions, whose matrix elements gives a system of linear equations. 
The key problem in this way is how to relate a such enlarged Hilbert space to the real physical states observed in high energy scattering.
Even after an explicit removal of ghost states, a physical interpretation is still not clear.

Since we did not assume any perturbative approximation or expansion, our calculation is valid for any value of the coupling constant $g$.
Indeed, we have shown that near $g\approx 0$ the theory can be quantized well.
This means that nonperturbative aspects of lightcone QCD should arise somewhere.
Recall that any lightcone theory has the trivial vacuum state.
So, in a general case potential troubles is moved to a very complicated Hamiltonian and to a quantization procedure.
We just conjecture that the observed obstructions is not technical but a genuine feature of the lightcone theory.
It is suitable to remind here about the lightcone quantization of the bosonic string theory where the critical dimension 
is given by a cumbersome check of the commutation relations of the Lorentz group.
In lightcone QCD we also see the drastic growing of complexity when we increase the dimension of space-time from three to four.
Namely, it was argued in Ref. \cite{avp_Hqcd} that at three space-time dimensions the residual gauge freedom can be used to make
the antisymmetric boundary condition for the field $A_i$ and the Hamiltonian contains a finite number of terms.

\section*{Acknowledgments}
This work was supported by RFFI 11-02-01395-a grant.

\end{document}